\documentstyle[aps,prd,floats,epsf,graphicx,amsfonts,colordvi]{revtex}

\begin{document}

\twocolumn[\hsize\textwidth\columnwidth\hsize\csname
@twocolumnfalse\endcsname

Ulm--TP/99--1

March 1999

\vspace{-25pt}

\preprint{Ulm--TP/99--1}

\draft
\title{
The Fluctuations of the Cosmic Microwave Background\\
for a Compact Hyperbolic Universe
}
\author{R.\,Aurich}

\address{
   Abteilung Theoretische Physik, Universit\"at Ulm\\
   Albert-Einstein-Allee 11, D-89069 Ulm \\
   Federal Republic of Germany\\
}

\date{\today}
\maketitle
\begin{abstract}
The fluctuations of the cosmic microwave background (CMB)
are investigated for a small open universe, i.\,e., one which is periodically
composed of a small fundamental cell.
The evolution of initial metric perturbations is computed using the
first 749 eigenmodes of the fundamental cell in the framework of
linear perturbation theory using a mixture of radiation and matter.
The fluctuations of the CMB are investigated for various density parameters
$\Omega_0$ taking into account the full Sachs-Wolfe effect.
The corresponding angular power spectrum $C_l$ is compared with recent
experiments.
\end{abstract}


\vskip2pc]

\narrowtext

\section{Introduction}

In recent years much attention has been paid to the possibility
that an open universe could have a non-trivial topology.
It is supposed that the 4-dimensional space-time can be represented
as the direct product ${\mathbb{R}} \times {\cal M}$,
where ${\cal M}$ is a compact three-manifold and the real line ${\mathbb{R}}$
represents time, and the space-time has a local structure as
in the Friedmann-Lema\^{\i}tre universe.
The topology is at least constrained, but not fixed by Einstein's theory of
gravitation because Einstein's equations deal as partial differential
equations with the local geometry of space-time, whereas the global structure
of space depends on the metric as well as on the topology.
A non-trivial topo\-logy can lead to a spatial closure in the universe
because of the global connectivity instead of a positive spatial curvature.

In order to manifest in observations, the topology should lead to a
small universe
\cite{EllSch86},
i.\,e., one in which light has been circled
round the universe at least once at our present epoch.
Such a small universe can be obtained from the universal covering space
by suitable identifications under a discrete group of isometries.
In the case of a flat universe the full isometry group
${\mathbb{R}}^3 \times \hbox{SO}(3)$ leads to 17 multi-connected types of
locally Euclidean spaces from which 10 are spatially compact, see e.\,g.\
\cite{LacReyLum95}.
The simplest case is a flat universe with a toroidal structure,
i.\,e., the topo\-logy of a three-torus.
For such rectangular fundamental cells the implications for the universe
are studied in
\cite{Sok93,Sta93,SteScoSil93,CosSmo95,CosSmoSta96}
yielding the conclusion that the dimension of the smallest toroidal structure
must be at least of the order of the horizon size to be compatible with
the COBE-DMR data
\cite{Ben96}.
The other compact, orientable flat spaces are investigated in \cite{LevScaSil98}
showing that periodicities with half the horizon size are marginally
consistent with the data.
In the case of positive curvature the elliptic topology is studied in e.\,g.\
\cite{PetSal68,Sol69}.
However, since astrophysical data suggest that the cosmological
density parameter $\Omega_0$ is sub-critical, see for example \cite{Pee98}, 
in the following only the case of negative curvature is considered.
There is now evidence for a non-vanishing cosmological constant $\Lambda$,
but this case will be considered in a forthcoming publication.
In this paper a vanishing cosmological constant $\Lambda=0$ is assumed.

For negative curvature there exists an infinite variety of possible
fundamental cells, see e.\,g.\
\cite{LacReyLum95,Thu79}
or the so-called census of compact hyperbolic manifolds of the
Geometry Centre at the University of Minnesota
\cite{SnapPea}.
In contrast to the Euclidean case there is no scaling freedom for
the fundamental cells.
Due to the rigidity theorem of Mostow
\cite{Mos73}
the volume as well as the lengths of closed geodesics are topological
invariants for a given hyperbolic 3-manifold.
Thus if the curvature scale, i.\,e., the density parameter $\Omega_0$
and $\Lambda$,
is given then the geometrical properties of ${\cal M}$ are fixed.

One manifestation of the non-trivial topology is given by multiple images
of objects at sufficient distances determined by the discrete group
of isometries $\Gamma$.
However, for a realistic value of $\Omega_0 \simeq 0.2\dots0.3$ the distance
to the nearest mirror images is larger than the range of
survey galaxy catalogs of roughly 200\dots600 Mpc.
Quasars are more distant but the quasar phenomenon is probably too
short-lived in order to observe multiple quasar images
because the distances and thus the look-back times of the images are
all distinct, in general (see \cite{LehLumUza98} and references therein).
Another manifestation of the topology is provided by pairs of circles of
correlated microwave radiation coming from the surface of last scattering
(SLS)
\cite{CorSpeSta98a,Wee98}.
The identical temperatures at points identified according to $\Gamma$
originate at the SLS having a red-shift of roughly $z\sim 1200$.
Unfortunately, in the case of $\Omega_0 \simeq 0.2\dots0.3$, the main
contribution of the CMB does not come from the SLS,
which corresponds to the naive Sachs-Wolfe effect
\cite{SacWol67},
but instead, from much nearer regions $z\ll 1200$
corresponding to the integrated Sachs-Wolfe effect, see e.\,g.\
\cite{GouSugSas91,KamSpe94,CorSpeSta98b} and below.
In the case of a flat universe where the integrated Sachs-Wolfe effect
is absent and the CMB is due to the SLS, the pairs of circles allow
the determination of the isometry group by determining the generators
of the group $\Gamma$.
For $\Omega_0 < 1$ one has to compute CMB for given examples of compact
models to obtain ideas of the expected structure of the CMB.
For two compact models the expected CMB is computed using the method of
images in \cite{BonPogSou98}.
This method requires the computation of the group elements of $\Gamma$
which is not an easy task since these groups are not free, i.\,e.,
there are relations among the generators of the group such that not
all products of generators yield new group elements.
Since the method of images requires only the distinct group elements
much attention has to be paid.
Nevertheless, the result of \cite{BonPogSou98}
is that only for $\Omega_0 \simeq 0.8$ a CMB is obtained
which is in accord with the COBE measurements.
An alternative to calculate the CMB demands the computation of the eigenmodes
of the considered 3-manifold.
In the case of the so-called Thurston manifold the first 14 eigenmodes
are computed using the boundary element method in \cite{Inoue98} and the
statistic of expansion coefficients of the eigenmodes is investigated
showing pseudo-random behaviour, where the term ``pseudo'' reflects the fact
that the coefficients are determined by the eigenmodes and not by
a genuine random process.

Interestingly, the possible volumes of compact hyperbolic 3-manifolds are
bounded from below, which means that there exists a hyperbolic 3-manifold
with minimal volume.
It is suggested that the creation probability of the universe increases
dramatically with decreasing volume, e.\,g.\ \cite{AtkPag82}.
Thus, from a cosmological point of view the most interesting hyperbolic
3-manifolds are those with volumes near to the volume of the smallest
hyperbolic 3-manifold.
Unfortunately, the smallest hyperbolic 3-manifold is unknown.
The two smallest known ones have volumes
$\hbox{vol}{\cal M}\simeq 0.98139 R^3$
\cite{Thu82}
and $\hbox{vol}{\cal M}\simeq 0.94272 R^3$
\cite{Wee85,MatFom88},
respectively,
where $R$ is the curvature radius of the universal covering space.
Another possibility is to consider not only manifolds but instead allowing
also orbifolds as possible models for the universe.
The difference is that orbifolds can possess points which are not
locally looking like the usual ${\mathbb{R}}^3$.
Orbifolds can possess rotation elements in their group of isometries.
Around the axis of a given rotation element the space has to be identified
with respect to the discrete angle of the rotation element which does not
happen in the usual ${\mathbb{R}}^3$.
However, as long as there is no elaborated quantum cosmology,
which describes the way in which the topology and topological defects
of the universe develop,
orbi\-folds are as good as mani\-folds for a model of the cosmos.
In this paper an orbi\-fold with volume
$\hbox{vol}{\cal M}\simeq 0.7173068 R^3$ is chosen.
For this system the fluctuations of the CMB are computed using the eigenmodes.
The initial scalar metric perturbations are expanded in terms of the
eigenmodes of the orbi\-fold such that the modes develop independently in the
framework of linear perturbation theory assuming adiabatic evolution.
From the metric perturbations the fluctuations of the CMB can then
be computed by the Sachs-Wolfe effect
\cite{SacWol67}.

The next section describes the selected orbifold.
Section \ref{computation_of_CMB} outlines the procedure for the
computation of the CMB.
The last section describes the properties of the CMB depending on
the density parameter $\Omega_0$.
Finally, the angular power spectrum $C_l$ of the fluctuations is
compared with experimental data from COBE, Saskatoon and QMAP.

\section{The geometric model}

The orbifold used in this paper is obtained from a Kleinian group
which yields a pentahedron as a fundamental cell
which in turn is symmetric along an intersection plane.
The pentahedron is divided by this intersection plane into two equal tetrahedra.
Thus the eigenmodes can be computed by desymmetrizing the pentahedron
(for more details, see \cite{AurMar96}).
A group-symmetry consideration shows that the eigenmodes of the pentahedron
obeying periodic boundary conditions decompose into two symmetry classes,
one having Dirichlet boundary conditions, i.\,e., $\psi=0$,
at the surface of the tetrahedron,
and the other having Neumann boundary conditions,
i.\,e., a vanishing normal derivative $\partial \psi/\partial \vec n =0$.
Using the tetrahedron with Dirichlet and Neumann boundary conditions
facilitates the numerical computation of the eigenmodes.
In the nomenclature of
\cite{Lan50,MacRei89,Mac96}
this tetrahedron is called $T_8$.
It has a volume $\hbox{vol}{\cal M}\simeq 0.3586534 R^3$ and is defined by
the dihedral angles
$$
\angle BC = \frac\pi 2 \; , \; \; 
\angle CA = \frac\pi 3 \; , \; \;
\angle AB = \frac\pi 4 \; ,
$$
$$
\angle DA = \frac\pi 2 \; , \; \;
\angle DB = \frac\pi 3 \; , \; \;
\angle DC = \frac\pi 5 \; ,
$$
where $A$, $B$, $C$ and $D$ are the four corner points.
For the tetrahedron $T_8$ the first 749 eigenmodes corresponding to
Dirichlet boundary conditions have been computed using the
boundary element method as described in \cite{AurMar96}.
It is worthwhile to note that there are only nine compact tetrahedra
in hyperbolic space and that $T_8$ is the only compact tetrahedron
whose generating group is not arithmetic
\cite{MacRei89}.
Furthermore the smallest tetrahedron, $T_3$, has a volume
$\hbox{vol}{\cal M}\simeq 0.03588506 R^3$ roughly ten times
smaller than the volume of $T_8$.

The CMB depends on the position of the observer within the fundamental cell.
The computations are carried out in the so-called upper half space
$$
{\cal H}_3 \; = \; \{ (x,y,z) \in {\mathbb{R}}^3 | z > 0 \}
$$
endowed with the hyperbolic metric
$$
ds^2 \; = \; \frac 1{z^2}\, (dx^2 + dy^2 + dz^2)
\hspace{10pt} ,
$$
yielding constant curvature $-1$.
In this model for hyperbolic space the tetrahedral cell is oriented
such that the corner points are approximately at
$A \simeq (0.4348,0,0.2537)$,
$B \simeq (0.3978,0.6889,0.2869)$,
$C \simeq (0,0,0.2824)$ and
$D \simeq (0,0,2.3829)$.
The observer is situated at $(0.15,0.2,0.5)$ which lies well within
the fundamental cell.

\section{The computation of the CMB}

\label{computation_of_CMB}

In order to compute the evolution of the initial scalar metric perturbations
one has to define a background model which describes the
homogeneous, isotropic space-time without any perturbations.
The perturbations are assumed to be sufficiently small such that the linear
perturbation theory is applicable.
Let us set the speed of light equal to $c=1$ and define the conformal time
$\eta$ by $a\, d\eta = dt$, where $a$ is the scale factor.
The background metric is chosen to be the Friedmann-Lema\^{\i}tre-Robertson-Walker metric
$$
ds^2 \; = \; a^2(\eta) \, ( d\eta^2 - \gamma_{ij}dx^i dx^j)
$$
with
$$
\gamma_{ij} \; = \; \delta_{ij} \, \left( 1 - \frac 14(x^2+y^2+z^2)
\right)^{-2}
$$
for the case of negative curvature.
The spatial part $\gamma_{ij}$ corresponds to the unit-ball model used
in hyperbolic geometry with the difference that the ball has here a radius
of two instead of one.
The Einstein equations, expressed in conformal time $\eta$, reduce for
the background metric to the time-time equation
\begin{equation}
\label{Einstein_0-0}
{a'}^2 - a^2 \; = \; \frac{8\pi G}3 T_0^0 a^4
\end{equation}
and to the trace of the Einstein equations
\begin{equation}
\label{Einstein_trace}
a'' - a \; = \; \frac{4\pi G}3 T_\mu^\mu a^3
\hspace{10pt} ,
\end{equation}
where $a' := da/d\eta$ and $T_\nu^\mu$ is the energy-momentum tensor.
$G$ denotes Newton's gravitational constant.
In the following a model with conventional relativistic hydrodynamic
matter behaving as a perfect fluid is assumed.
The energy-momentum tensor, which is then diagonal,
is described in terms of the energy density $\varepsilon$,
the pressure $p$ and the 4-velocity $u^\mu$ as
$$
T_\nu^\mu \; = \; (\varepsilon+p) u^\mu u_\nu - p\, \delta_\nu^\mu
\hspace{10pt} .
$$
Assuming a two-component model containing radiation with energy density
$\varepsilon_r \propto a^{-4}$ and cold dark matter with energy density
$\varepsilon_m \propto a^{-3}$, the pressure perturbation $\delta p$
is for a vanishing entropy perturbation $\delta S=0$ given by
$$
\delta p \; = \;
\left.\left(\frac{\partial p}{\partial\varepsilon}\right)\right|_S \,
\delta\varepsilon \; =: \; c_s^2 \delta\varepsilon
$$
where $c_s$ can be interpreted as the sound velocity.
From this follows with $\varepsilon = \varepsilon_m+\varepsilon_r$
and $p = \frac 13 \varepsilon_r$, see e.\,g.\ \cite{MukFelBra92},
$$
c_s^2 \; = \; \frac 1{3 + \frac 94 \varepsilon_m/\varepsilon_r}
\hspace{10pt} .
$$
For this two-component model the equations 
(\ref{Einstein_0-0}) and (\ref{Einstein_trace}) are solved by
$$
a(\eta) \; = \; \frac{2 a_{\hbox{\scriptsize{eq}}}}{\hat \eta^2} \, \left\{
\hat \eta \sinh \eta + \cosh \eta - 1 \right\}
\; \; \; ,
$$
$$
\hat \eta \; := \;
\sqrt{\frac{2 a_{\hbox{\scriptsize{eq}}}^2}{2\pi G \varepsilon_r a^4}}\; ,
$$
where $a_{\hbox{\scriptsize{eq}}}$ is the scale factor at the time of
equal matter and radiation density, i.\,e., $\varepsilon_m=\varepsilon_r$.
In the gauge-invariant formalism
\cite{Bar80,KodSas84,MukFelBra92}
the perturbed metric can be written as
$$
ds^2 \; = \; a^2(\eta) \left\{ (1+2\Phi) d\eta^2 -
(1-2\Psi) \gamma_{ij} dx^i dx^j \right\}
\hspace{10pt} .
$$
For a diagonal energy-momentum tensor as assumed above
one gets $\Phi = \Psi$ which can be considered as a generalized
Newtonian gravitational potential.
Assuming vanishing entropy perturbations $\delta S=0$,
the gauge-invariant formalism for the evolution of the
metric perturbation $\Phi$ gives in first order perturbation theory
\cite{Bar80,KodSas84,MukFelBra92}
\begin{eqnarray}\nonumber
\Phi'' & + & 3 \hat H (1+c_s^2) \Phi' \, - \, c_s^2 \Delta \Phi
\\ & &
\label{metric_perturbation} \hspace{-10pt}
+ \, \{2 \hat H' + (1+3c_s^2)(\hat H+1)\} \Phi \; = \; 0
\hspace{10pt} . \hspace{10pt}
\end{eqnarray}
The prime denotes differentiation with respect to $\eta$ and
$\hat H := a'/a$.
The Laplace-Beltrami operator of the hyperbolic space is denoted
by $\Delta$.
Expanding the metric perturbation $\Phi$
with respect to the eigenmodes $\psi_n(\vec x\,)$ of the orbifold, i.\,e.,
$$
\Phi(\eta,\vec x\,) \; = \; \sum_{n=1}^\infty f_n(\eta)\, \psi_n(\vec x\,)
\hspace{10pt} ,
$$
yields for $f_n(\eta)$ the differential equation
\begin{eqnarray}\nonumber
f_n''(\eta) & + & 3 \hat H (1+c_s^2) f_n'(\eta) 
\\ & & \hspace{-25pt} \label{deq_f}
+ \, \{ c_s^2 E_n + 2 \hat H' + (1+3c_s^2)(\hat H+1)\} f_n(\eta) \; = \; 0
\hspace{5pt} . \hspace{5pt}
\end{eqnarray}
Here $E_n$ denotes the eigenvalue corresponding to $\psi_n$,
i.\,e., $(\Delta+E_n) \psi_n = 0$ with Dirichlet and Neumann boundary
conditions, respectively.
The wavenumber is given by $k_n = \sqrt{E_n-1}$.
The computation of the time evolution of $\Phi(\eta,\vec x\,)$ is
now reduced to the integration of the ordinary differential equation
(\ref{deq_f}) which can easily be done numerically.
It is worthwhile to note that the expansion runs only over discrete levels.
At this point enters the compact nature of the model of the universe
assumed here.

Now one is left to define the initial conditions of $f_n(\eta_i)$
for some initial time $\eta_i$.
Inflationary models suggest a scale invariant, so-called
Harrison-Zel'do\-vitch spectrum for the density perturbation modes $\delta_k$
up to logarithmic corrections, see \cite{MukFelBra92} and references
therein.
Since the corrections depend on the details of the inflationary model,
we assume here
\begin{equation}
\label{ansatz}
f_n(\eta_i) \; = \; \frac{\alpha}{k_n^{3/2}}
\hspace{10pt} \hbox{ and } \hspace{10pt}
f_n'(\eta_i) \; = \; 0
\hspace{10pt} ,
\end{equation}
which carries over to a Harrison-Zel'dovitch spectrum.
The constant $\alpha$ is fixed later such that the order of the fluctuations
$\delta T/T$ are in agreement with the COBE results.
The value of $\eta_i=0.001$ is used which is well within the radiation
dominated epoch such that the transition to the matter dominated epoch is
included in the computations.
All Dirichlet eigenmodes with $k_n < k_{\hbox{\scriptsize{max}}} = 55$,
i.\,e., with $E_n<3026$, are taken into account
$(n_{\hbox{\scriptsize{max}}}=749)$.

It is worthwhile to emphasize that the initial coefficients $f_n(\eta_i)$ are
given by (\ref{ansatz}).
Thus, $f_n(\eta_i)$ is not considered as random variable which could be, e.\,g.,
distributed as a Gaussian.
Therefore, the randomness in $\Phi$ is solely due to the properties of the
eigenmodes of the considered orbifold.
The eigenmodes are square-normalized with respect to the volume of the
orbifold using the hyperbolic metric.
The only freedom is a factor $\pm 1$ where that sign is used which comes out
from the boundary-element calculations.

From the metric perturbations $\Phi$ the temperature fluctuations $\delta T/T$
are computed by the Sachs-Wolfe effect
\cite{SacWol67}
\begin{eqnarray}
\frac{\delta T}T & = & \nonumber \frac 13
\Phi(\eta_{\hbox{\scriptsize{SLS}}},\vec x(\eta_{\hbox{\scriptsize{SLS}}}))
\\ & & \label{SWE} \; \; \; + \,
2 \, \int_{\vec x(\eta_{\hbox{\scriptsize{SLS}}})}^{\vec x(\eta_0)} d\eta
\frac{\partial\Phi(\eta,\vec x(\eta))}{\partial\eta}
\hspace{10pt} ,
\end{eqnarray}
where $\eta_{\hbox{\scriptsize{SLS}}}$ denotes the time of last scattering assumed
to be at $z\simeq 1200$, and $\eta_0$ is the present time.
The factor $1/3$ in the first term corresponding to the naive Sachs-Wolfe
effect (NSW) is justified in the models considered below since the time of
decoupling occurs well within the matter dominated epoch.
The second term is the integrated Sachs-Wolfe effect (ISW).
Note, that (\ref{SWE}) is valid only on scales large compared with the horizon
at the time $\eta_{\hbox{\scriptsize{SLS}}}$.
In table \ref{tab_SWE} the angle $\Theta_H$ under which the horizon appears
is shown together with the angle $\Theta_k$ which is the angle under which
the highest eigenmode fluctuation appears.
Since $\Theta_k$ is at least roughly two-times larger than $\Theta_H$,
equation (\ref{SWE}) can be used for all considered eigenmodes.

In a multi-component model, where the components possess different sound velocities,
the time evolution generates an entropy perturbation even if one starts with
$\delta S=0$
\cite{KodSas84}.
Then the right-hand side of (\ref{metric_perturbation}) is not strictly zero.
However, the corrections are very small because of the large number ratio of
photons to dark matter particles.
Furthermore, the corrections are negligible as long as the wavelengths of
the eigenmodes are larger than the horizon.
The highest considered eigenmode $(k_{\hbox{\scriptsize{max}}} = 55)$ enters
the horizon at the conformal time
$\eta=2\pi/k_{\hbox{\scriptsize{max}}} \simeq 0.11$ and all other eigenmodes
correspondingly later.
This is to be compared with $\eta_{\hbox{\scriptsize{SLS}}}$ which is
$\eta_{\hbox{\scriptsize{SLS}}}\simeq 0.063$ and
$\eta_{\hbox{\scriptsize{SLS}}}\simeq 0.033$
for $\Omega_0=0.2$ and $\Omega_0=0.6$, respectively.
Thus, the entropy perturbations can only slightly alter the ISW.

\section{Properties of the CMB}

\label{properties_of_CMB}

\begin{figure}[ttt]
\begin{center}
\includegraphics[width=8.5cm]{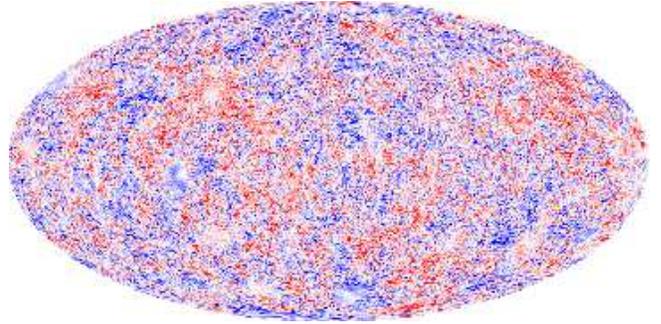}
\end{center}
\vspace*{-3pt}
\caption{\label{Fig:CMB_Mollweide_30}
CMB for $\Omega_0=0.3$ using the Mollweide projection.
}
\end{figure}

\begin{figure}[ttt]
\begin{center}
\includegraphics[width=8.5cm]{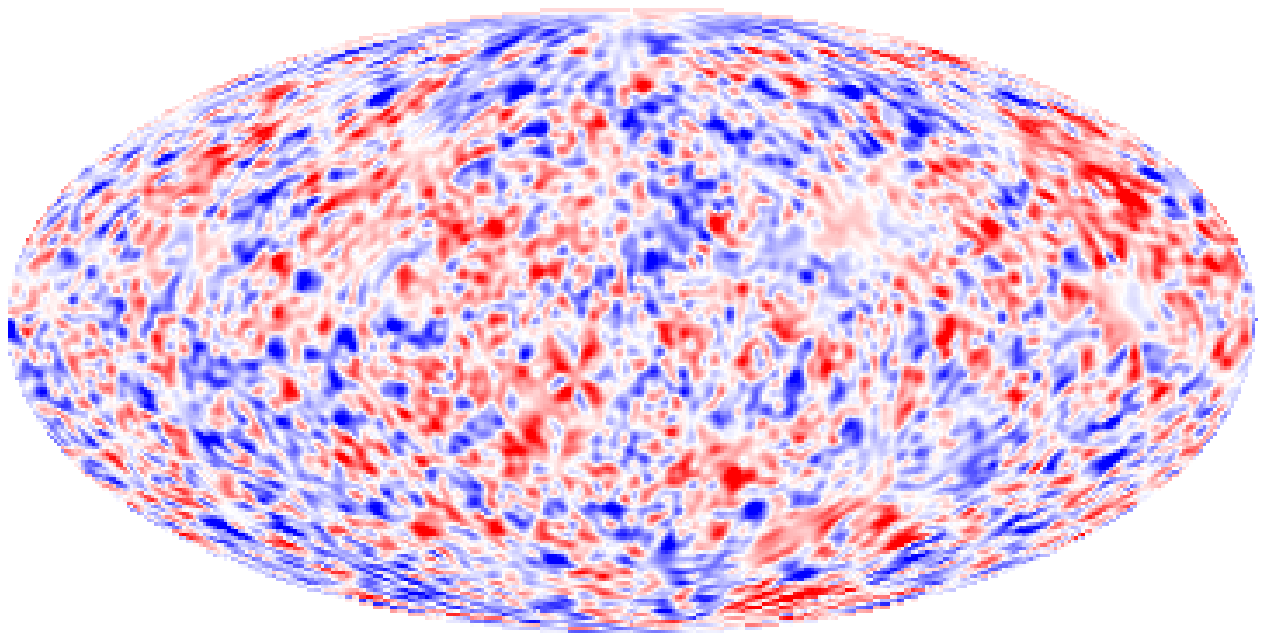}
\end{center}
\vspace*{-3pt}
\caption{\label{Fig:CMB_Mollweide_60}
CMB for $\Omega_0=0.6$ using the Mollweide projection.
}
\end{figure}

\begin{figure}[htb]
\begin{center}
\includegraphics[width=8.5cm]{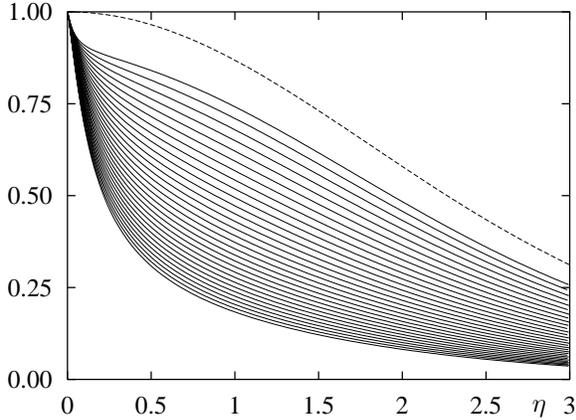}
\put(-20,12){$\eta$}
\end{center}
\vspace*{-3pt}
\caption{\label{Fig:f_n}
The dependence of $k_n^{3/2} f_n(\eta)$ on the eigenvalue $E_n$ for the case
$\Omega_0=0.3$ is shown, where $\eta_0=2.3746$ and $\alpha=1$.
The upper full curve corresponds to $E_n=100$ and the lowest one
to $E_n=3000$.
For the intermediate curves the energy is increased in steps of 100.
The dashed curve represents the result
$f_{\hbox{\scriptsize{mat}}}(\eta) = 5(\sinh^2\eta-3\eta\sinh\eta+4\cosh\eta-4)
/(\cosh\eta-1)^3$
belonging to a pure matter model with $c_s=0$ used in some related works.
}
\end{figure}

The CMB is computed for different densities $\Omega_0$ with respect to
the Hubble constant $h=0.6$ in units of
$100 \hbox{ km } \hbox{s}^{-1} \hbox{Mpc}^{-1}$.
The radiation density $\varepsilon_r$ is chosen in agreement with the
present background radiation temperature of $T=2.728\hbox{ K}$.
In figures \ref{Fig:CMB_Mollweide_30} and \ref{Fig:CMB_Mollweide_60}
the fluctuations of the CMB are shown for
$\Omega_0=0.3$ and $0.6$, respectively.
The monopole and the dipole contribution is subtracted such that the first
non-vanishing multipole is the quadrupole in accordance with the
usual representation of the CMB fluctuations.

A quantitative measure of the scale of the fluctuations
is provided by the angular power spectrum $C_l$ defined by
$$
C_l \; = \; \frac{1}{2l+1}\,\sum_{m=-l}^l |a_{lm}|^2
\hspace{10pt} ,
$$
where $a_{lm}$ are the expansion coefficients of $\delta T$ with respect
to the spherical harmonics $Y_l^m(\theta,\phi)$.
In the following the angular power spectrum $C_l$ is compared with
the data measured for $l<30$ by COBE
\cite{Teg96},
around $l\sim 100$ by QMAP
\cite{CosDevHer98}
and above $l\sim 80$ by the Saskatoon experiment
\cite{NetDevJar97}.
These experiments provide evidence that the angular power spectrum
increases up to a maximum around $l\simeq 200$.

The figures \ref{Fig:CMB_Mollweide_30} and \ref{Fig:CMB_Mollweide_60}
show fluctuations on finer scales with decreasing density $\Omega_0$.
This is due to the cut-off in the wavenumbers
$k_n < k_{\hbox{\scriptsize{max}}}$.
For a genuine Harrison-Zel'dovitch spectrum having no cut-off
there are fluctuations on all scales.
However, the amplitudes belonging to the different scales depend on
$\Omega_0$ by the specific form of decay of $f_n(\eta)$ via the ISW.
The decay of $f_n(\eta)$ is faster for smaller
densities $\Omega_0$ with increasing eigenvalue $E_n$ as shown in
figure \ref{Fig:f_n} for $\Omega_0=0.3$.
An estimate of the angular contributions which are maximally
taken into account in the angular power spectrum $C_l$ for a given
$k_{\hbox{\scriptsize{max}}}$ can be obtained as follows.
The smallest wavelength
$\lambda_{\hbox{\scriptsize{min}}}\simeq\frac{2\pi}
{k_{\hbox{\scriptsize{max}}}}$
of the eigenmodes determines the finest scale of the fluctuations on the SLS.
The angle $\Theta_k$ under which the smallest scale fluctuations are seen,
is given by
$$
\tan\frac{\Theta_k}2 \; = \; \frac{\tanh \frac{\lambda_{\hbox{\scriptsize{min}}}}2}
{\sinh(\eta_0-\eta_{\hbox{\scriptsize{SLS}}})}
\hspace{10pt} .
$$
Since the $l$th spherical harmonics has along the ``circumference'' $2l$ zeros
in $[-180^\circ,180^\circ]$ one gets the rule of thumb
that a given $\Theta$ corresponds roughly to $l \simeq 180^\circ/\Theta$.
In table \ref{tab_SWE} the distance $\eta_0-\eta_{\hbox{\scriptsize{SLS}}}$,
the angle $\Theta_k$ under which $\lambda_{\hbox{\scriptsize{min}}}$ appears
and the corresponding angular contribution
$l_k := 180^\circ/\Theta_k$ are given for several values of $\Omega_0$.

\begin{figure}[ttt]
\begin{center}
\hspace*{-15pt}\includegraphics[width=9cm]{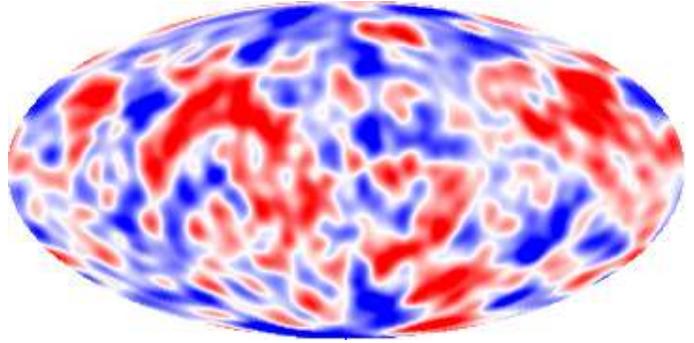}
\end{center}
\caption{\label{Fig:CMB_Mollweide_smooth_30}
A Gaussian smoothing of figure \ref{Fig:CMB_Mollweide_30},
i.\,e., the case $\Omega_0=0.3$, is shown with a resolution of $10^\circ$.
}
\end{figure}

To get an impression of the fluctuations as seen with the $10^\circ$
resolution of COBE, a Gaussian smoothing of figure \ref{Fig:CMB_Mollweide_30},
i.\,e., for $\Omega_0=0.3$, is presented in
figure \ref{Fig:CMB_Mollweide_smooth_30} with that resolution.

\begin{figure}[htb]
\begin{center}
\hspace*{-15pt}\includegraphics[width=9cm]{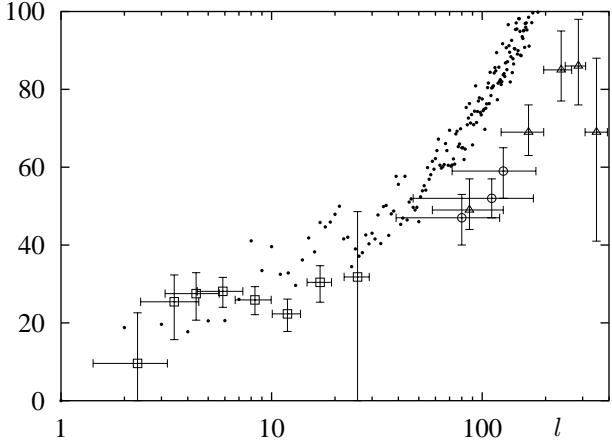}
\put(-27,12){$l$}
\end{center}
\vspace*{-15pt}
\caption{\label{Fig:CMB_aps_20}
The angular power spectrum $\sqrt{l(l+1) C_l /2\pi}$
for $\Omega_0=0.2$ for the hyperbolic model (full dots) in comparison with
the COBE $(\square)$, Saskatoon $(\triangle)$ and QMAP $(\bigcirc)$ data.
}
\end{figure}

\begin{figure}[htb]
\begin{center}
\hspace*{-15pt}\includegraphics[width=9cm]{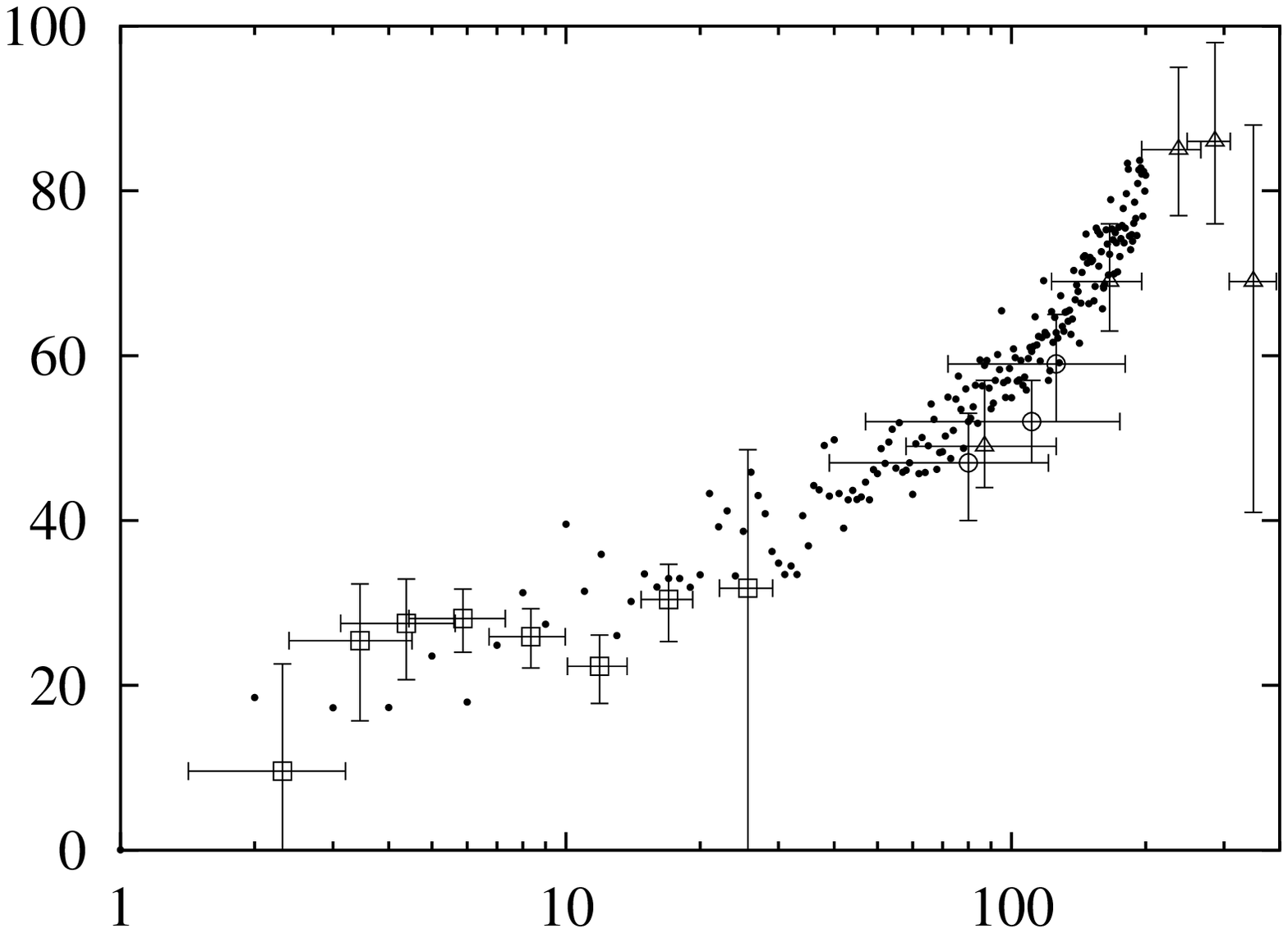}
\put(-27,12){$l$}
\end{center}
\vspace*{-15pt}
\caption{\label{Fig:CMB_aps_30}
The angular power spectrum $\sqrt{l(l+1) C_l /2\pi}$
for $\Omega_0=0.3$ with the same symbols as in figure
\protect\ref{Fig:CMB_aps_20}.
}
\end{figure}

\begin{figure}[htb]
\begin{center}
\hspace*{-15pt}\includegraphics[width=9cm]{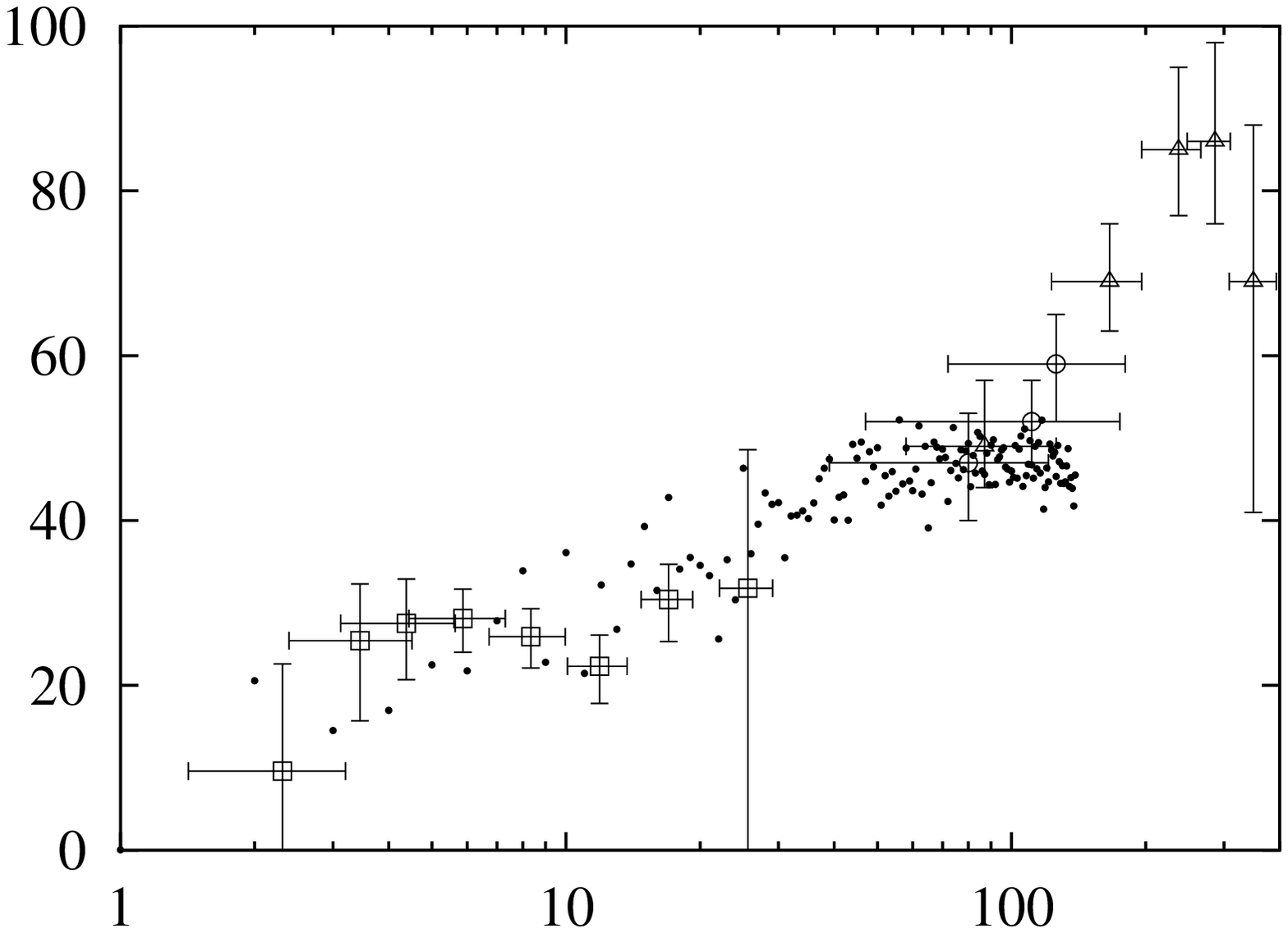}
\put(-27,12){$l$}
\end{center}
\vspace*{-15pt}
\caption{\label{Fig:CMB_aps_40}
The angular power spectrum $\sqrt{l(l+1) C_l /2\pi}$
for $\Omega_0=0.4$ with the same symbols as in figure
\protect\ref{Fig:CMB_aps_20}.
}
\end{figure}

\begin{figure}[htb]
\begin{center}
\hspace*{-15pt}\includegraphics[width=9cm]{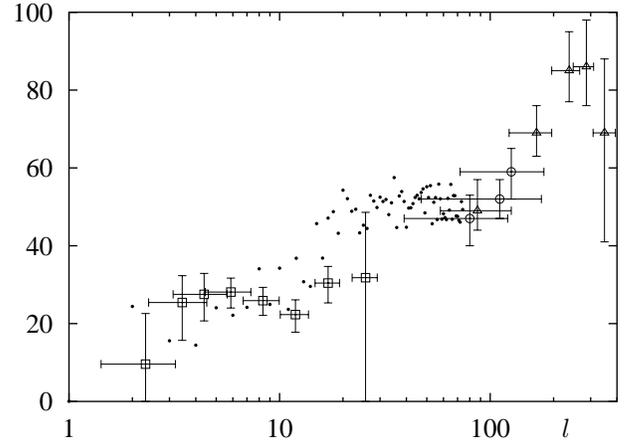}
\put(-27,12){$l$}
\end{center}
\vspace*{-15pt}
\caption{\label{Fig:CMB_aps_60}
The angular power spectrum $\sqrt{l(l+1) C_l /2\pi}$
for $\Omega_0=0.6$ with the same symbols as in figure
\protect\ref{Fig:CMB_aps_20}.
}
\end{figure}

The figures \ref{Fig:CMB_aps_20} to \ref{Fig:CMB_aps_60} show
the angular power spectrum for $\Omega_0=0.2, 0.3, 0.4$ and $0.6$,
where the abscissa shows $\sqrt{l(l+1) C_l /2\pi}$ in $\mu \hbox{K}$.
The data for the compact hyperbolic models are shown up to roughly $l_k$.
A reasonable agreement is observed for $\Omega_0 \simeq 0.3\dots 0.4$.
In the case $\Omega_0=0.2$ $C_l$ increases too fast with increasing
$l$ in comparison with the experimental data.
For $\Omega_0=0.4$ one observes a saturation above $l\simeq 40$.
Then the further increase of the $C_l$ has to come from processes,
like acoustic oscillations, getting important on scales of the horizon size
at the SLS around $l_H=200 \dots 300$, see table \ref{tab_SWE}.
The necessary contributions are not considered here but the results obtained
from simply-connected models should then apply since the corresponding scales
are small in comparison with the size of the fundamental cell.
Further effects like the reionization, gravitational lensing and
the Sunyaev-Zel'dovitch effect influence the fluctuations on a scale of
order $\Theta \simeq 1^\circ$ and
thus the values of $C_l$ for correspondingly large $l$.
More important are the low multipoles since they have ruled out
a toroidal structure in the case of a flat universe,
at least for periodicities significantly below the horizon size
\cite{SteScoSil93,CosSmo95,LevScaSil98}.
In the flat case the first multipoles were too small in comparison
with the multipoles around $l\simeq 20$.
Such an effect is absent in the hyperbolic case and thus
a ``small'' universe is not ruled out.
This result is at variance with \cite{BonPogSou98}
where agreement with COBE is only obtained for very high densities
$\Omega_0\simeq 0.8$.

\begin{figure}[htb]
\begin{center}
\hspace*{-15pt}\includegraphics[width=9cm]{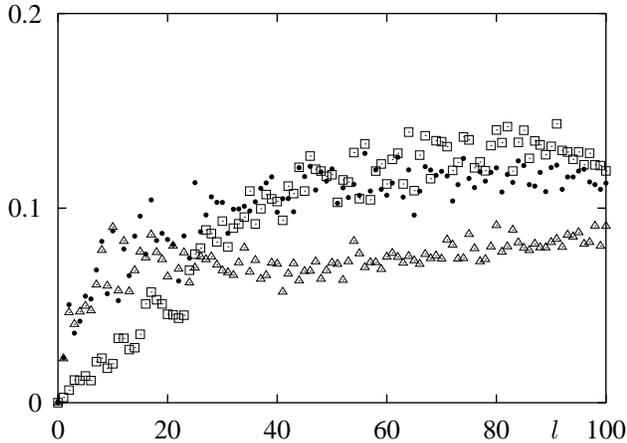}
\put(-27,12){$l$}
\end{center}
\vspace*{-15pt}
\caption{\label{Fig:CMB_ISW_NSW_40}
For $\Omega_0=0.4$ the individual contributions of the NSW $(\square)$
and the ISW $(\triangle)$ to the angular power spectrum
$\sqrt{l(l+1) C_l /2\pi}$ are shown
as well as their combined effect (full dots) $(\alpha = 1)$.
}
\end{figure}

In order to show the increasing significance of the ISW
with decreasing $\Omega_0$, table \ref{tab_SWE} shows the rms
of the two terms in (\ref{SWE}) using $\alpha=1$ in (\ref{ansatz}).
The rms value of the NSW is nearly constant
because the fluctuations are determined by (\ref{ansatz}),
i.\,e., by $\alpha$ independently of the distance to the SLS.
In contrast the ISW diminishes towards $\Omega_0=1$.
For $\Omega_0=1$ no ISW contribution occurs because of
$d\Phi/d\eta=0$ in this case.
The figure \ref{Fig:CMB_ISW_NSW_40} shows the contributions to
the $C_l$ spectrum separately for the NSW and for the ISW
as well as both contributions together as observed in nature.
The case $\Omega_0=0.4$ is shown with $\alpha=1$.
The importance of the eigenmodes with respect to the contribution to the ISW
is determined by two competing effects.
On the one hand higher eigenmodes are oscillating faster such that their
contribution to the integral is less important.
However, also the derivative $f_n'(\eta)$ determines the significance,
and it is this derivative which increases with increasing eigenvalues as can
be inferred from figure \ref{Fig:f_n}.
Numerically both effects seem to cancel such that the ISW is nearly constant
with respect to $C_l$.
At $\Omega_0=0.3$ the NSW and the ISW are of equal significance and 
for less values of $\Omega_0$ the ISW dominates.
This can lead to difficulties in detecting paired circles.

\begin{figure}[htb]
\begin{center}
\hspace*{-15pt}\includegraphics[width=9cm]{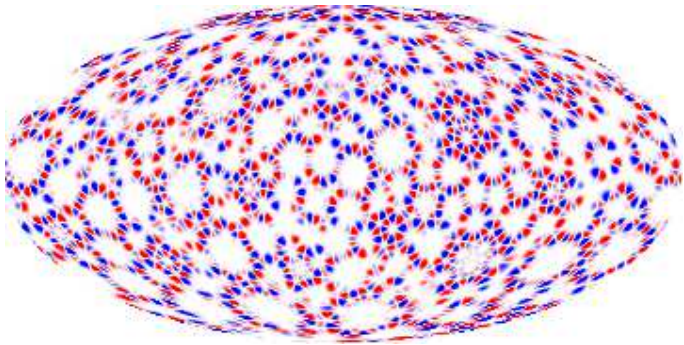}
\end{center}
\caption{\label{Fig:CMB_NSW_E93}
The NSW contribution of the lowest eigenmode $(E_1\simeq 93.11)$ to the
CMB is shown for the case $\Omega_0=0.2$.
}
\end{figure}

\begin{figure}[htb]
\begin{center}
\hspace*{-15pt}\includegraphics[width=9cm]{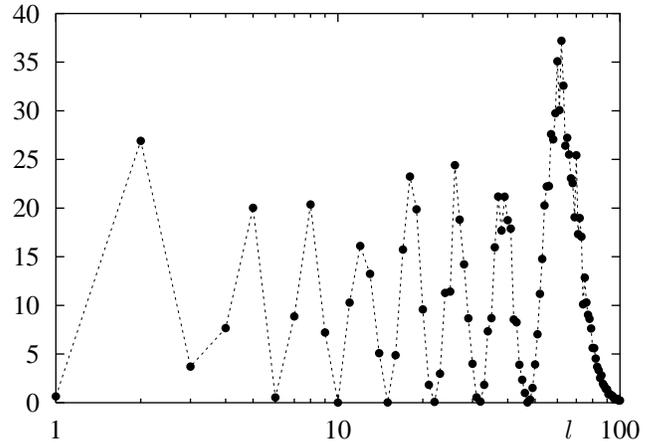}
\put(-27,12){$l$}
\end{center}
\caption{\label{Fig:Cl_NSW_E93}
The $C_l$ spectrum is shown for the NSW contribution of figure
\protect\ref{Fig:CMB_NSW_E93}.
Here $C_l$ is shown instead of $\sqrt{l(l+1) C_l /2\pi}$ as in the other figures.
The scaling at the abscissa is arbitrarily.
}
\end{figure}

It is interesting to note that the large scale power is not only caused
by the lowest eigenmodes but also by the large scale structure generated
by the isometry group describing the fundamental cell.
To stress that fact figure \ref{Fig:CMB_NSW_E93} shows the NSW contribution
of the lowest eigenmode having an eigenvalue $E_1\simeq 93.11$ which has only
one maximum within the tetrahedral cell.
One clearly observes large circles of different radii having a scale larger
than the wavelength of the eigenmode.
The $C_l$ spectrum, shown in figure \ref{Fig:Cl_NSW_E93}, has a periodic
structure with different maxima corresponding to the circles of different radii
and to the structure formed by the arrangement of the circles itself.
Only the wide last maximum around $l=60$ corresponds to the peak expected
from the wavelength of the lowest eigenmode.
If the peaks due to the circles are pronounced enough in comparison with the
ISW, there could survive a periodic structure in the full $C_l$ spectrum
yielding some information about the isometry group.

In conclusion small hyperbolic universes are not ruled out
in contrast to the flat case.
A reasonable agreement with the experimental data is provided by the
hyperbolic models with densities around $\Omega_0\simeq 0.3\dots 0.4$.
In this work an orbifold instead of a manifold is investigated,
however, the statistical properties of the eigenmodes are expected to be
the same for orbifolds and manifolds.
Thus, since the volume of the considered pentahedron is of the same
order as of the Weeks and Thurston manifolds,
the comoving wavenumbers $k_n$ of these models are comparable
since their mean behaviour is determined by Weyl's law.
(Weyl's law, which counts the mean number of eigenvalues below a given energy,
is derived in \cite{AurMar96} for general orbifolds.)
In these models no supercurvature modes are expected
which are absent in the considered pentahedron and could, if present,
alter the angular power spectrum.
The individual details of the sky maps would differ between the models,
but the angular power spectra should show the same behaviour.
However, the details of the maps already depend on the
locations of the observer within the given fundamental cell.
After all, such a small universe has a uniform CMB because in all directions
the fluctuations of the same fundamental cell contribute to the CMB
and thus circumvents the horizon problem.
In addition, the Machian paradox is solved in a new manner by the
non-trivial topology.
A future work will include the cosmological constant $\Lambda$ in order
to investigate its effect on the structure of the CMB.

\acknowledgments

I would like to thank the HLRZ at J\"ulich, the HLRS at Stuttgart
and the Rechenzentrum of the University of Karlsruhe for the access
to their computers.
I also wish to thank N.\,Cornish, J.\,Levin and the unknown referee
for useful comments.



\onecolumn

\begin{table}[htb]\center
\begin{tabular}{|c|c|c|c|c|c|c|c|}
\tableline
$\Omega_0$ & $\Theta_H$ & $l_H$& $\eta_0-\eta_{\hbox{\scriptsize{SLS}}}\;$ &
$\Theta_k\;$ & $l_k\;$ &
rms of NSW$\;$  & rms of ISW$\;$ \\
\tableline
0.2  & 0.46$^\circ$ & 391 & 2.7527 & 0.8$^\circ$  & 215 &   0.066   &   0.106  \\
0.3  & 0.60$^\circ$ & 300 & 2.3214 & 1.3$^\circ$  & 139 &   0.066   &   0.079  \\
0.4  & 0.73$^\circ$ & 247 & 1.9863 & 1.8$^\circ$  &  98 &   0.067   &   0.064  \\
0.6  & 0.94$^\circ$ & 191 & 1.4409 & 3.3$^\circ$  &  55 &   0.070   &   0.044  \\
0.8  & 1.12$^\circ$ & 161 & 0.9322 & 6.1$^\circ$  &  30 &   0.069   &   0.026  \\
\tableline
\end{tabular}
\caption{ \label{tab_SWE}
The angle $\Theta_H$, under which the horizon at the SLS is seen,
the corresponding $l_H:=180^\circ/\Theta_H$,
the distance to the SLS, the angle $\Theta_k$ due to the cut-off
$k_{\hbox{\scriptsize{max}}}$ and the corresponding $l_k$ are shown
as well as the rms of the NSW and the ISW contribution, respectively,
in (\ref{SWE}) for $\alpha=1$.}
\end{table}

\end{document}